\documentclass[12pt,journal,final,onecolumn]{IEEEtran}

\usepackage[dvips]{graphicx}
\usepackage{epsfig}
\usepackage[cmex10]{amsmath}
\usepackage{amssymb}
\usepackage{amsthm}
\usepackage{amsfonts}
\usepackage{bm}
\usepackage{cite}
\usepackage[svgnames]{xcolor} 
\usepackage{pstricks,pst-node,pst-plot,pstricks-add}
\usepackage[tight,footnotesize]{subfigure}
\usepackage{algpseudocode}
\usepackage{algorithm}

\graphicspath{{figs/}}

\input{niesen.def}

\begin{document}

\title{Fundamental Limits of Caching}

\author{Mohammad Ali Maddah-Ali and  Urs Niesen%
\thanks{This paper was presented in part at the International Symposium
on Information Theory, July 2013.}%
\thanks{The authors are with Bell Labs, Alcatel-Lucent. Emails:
\{mohammadali.maddah-ali, urs.niesen\}@alcatel-lucent.com}%
}

\maketitle

\begin{abstract} 
    Caching is a technique to reduce peak traffic rates by prefetching
    popular content into memories at the end users. Conventionally,
    these memories are used to deliver requested content in part from a
    locally cached copy rather than through the network. The gain
    offered by this approach, which we term \emph{local caching gain},
    depends on the \emph{local} cache size (i.e, the memory available at
    each individual user). In this paper, we introduce and exploit a
    second, \emph{global}, caching gain not utilized by conventional
    caching schemes.  This gain depends on the aggregate \emph{global}
    cache size (i.e., the cumulative memory available at all users),
    even though there is no cooperation among the users. 

    To evaluate and isolate these two gains, we introduce an
    information-theoretic formulation of the caching problem focusing on
    its basic structure. For this setting, we propose a novel coded
    caching scheme that exploits both local and global caching gains,
    leading to a multiplicative improvement in the peak rate compared to
    previously known schemes.  In particular, the improvement can be on
    the order of the number of users in the network.  Moreover, we argue
    that the performance of the proposed scheme is within a constant
    factor of the information-theoretic optimum for all values of the
    problem parameters.  
\end{abstract}

\section{Introduction}
\label{sec:introduction}

The high temporal variability of network traffic results in
communication systems that are congested during peak-traffic times and
underutilized during off-peak times. One approach to reduce peak traffic
is to take advantage of memories distributed across the network (at
end users, servers, routers, \dots) to duplicate content.  This
duplication of content, called content placement or caching, is
performed during off-peak hours when network resources are abundant.
During peak hours, when network resources are scarce, user requests can
then be served from these caches, reducing network congestion. In this
manner, caching effectively allows to shift traffic from peak to
off-peak hours, thereby smoothing out traffic variability and reducing
congestion.

From the above discussion, we see that the caching problem consists of
two distinct phases. The first phase is the \emph{placement phase},
which is based solely on the statistics of the user demands. In this
phase, the network is not congested, and the main limitation is the size
of the cache memories. The second phase is the \emph{delivery phase},
which is performed once the actual demands of the users have been
revealed.  In this phase, the network is congested, and the main
limitation is the rate required to serve the requested content.

Various versions of this problem have been studied, with the focus being
mainly on exploiting the history or statistics of the user
demands~\cite{dowdy82,almeroth96,dan96,korupolu99,meyerson01,baev08,borst10}.
In these papers, the operation of the delivery phase is fixed to consist
of simple orthogonal unicast or multicast transmissions. Assuming this
method of delivery, the content placement is then optimized. The gain of
caching in this approach results from making popular content available
locally. In another line of research, the objective is to optimize the
delivery phase for fixed known cache contents and for specific
demands~\cite{birk06,bar-yossef11} (see also the discussion in
Section~\ref{sec:discussion_index}). 

As pointed out above, the gain from traditional uncoded caching
approaches derives from making content available locally: if a user
requests some content that is stored in its cache, this request can be
served from its local memory. We hence call this the \emph{local caching
gain}. This gain is relevant if the local cache memory is large enough
such that a sizable fraction of the total (popular) content can be
stored locally.  On the other hand, if the size of the local caches is
small compared to the total amount of content, then this gain is
insignificant. 

In this paper, we propose a novel coded caching approach that, in
addition to the local caching gain, is able to achieve a \emph{global
caching gain}.  This gain derives from \emph{jointly} optimizing both
the placement and delivery phases, ensuring that in the delivery phase
several \emph{different} demands can be satisfied with a single coded
multicast transmission. Since the content placement is performed without
knowledge of the actual demands, in order to achieve this gain the
placement phase must be carefully designed such that these multicasting
opportunities are created \emph{simultaneously} for all possible
requests in the delivery phase.  We show that this global caching gain
is relevant if the aggregate global cache size is large enough compared
to the total amount of content. Thus, even though the caches cannot
cooperate, the sum of the cache sizes becomes an important system
parameter.

\begin{figure}[htbp]
    \centering
    \hspace{0.12cm}\includegraphics{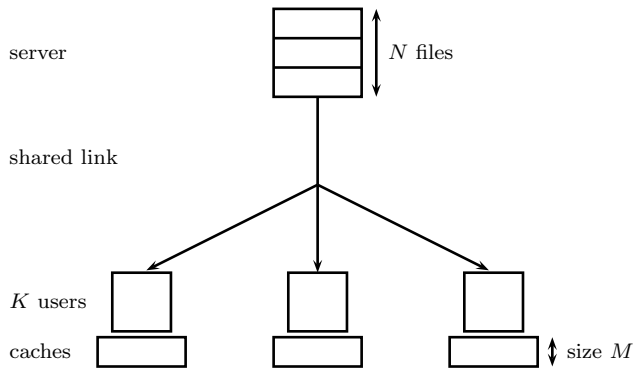}
    \caption{Caching system considered in this paper. A server
        containing $N$ files of size $F$ bits each is connected through
        a shared link to $K$ users each with an isolated cache of size
        $MF$ bits. The goal is to design the placement phase and
        the delivery phase such that the peak rate (i.e. the load
        normalized by the file size) of the shared bottleneck link is
        minimized. In the figure, $N=K=3$ and $M=1$.} 
    \label{fig:setting}
\end{figure}

To formally analyze the performance of the proposed coded caching
approach, and in order to evaluate and isolate these two gains, we
introduce a new, information-theoretic formulation of the caching
problem focusing on its basic structure. In our setting, depicted in
Fig.~\ref{fig:setting}, $K$ users are connected to a server through a
shared, error-free link. The server has a database of $N$ files of equal
size. Each of the users has access to a cache memory big enough to store
$M$ of the files. During the placement phase, the caches are filled as a
function of the database.  During the delivery phase, each user may ask
for any one of the $N$ possible files. The objective is to design the
placement and delivery phases such that the load of the shared link in
the delivery phase is minimized. For simplicity, we restrict the
discussion in the introduction section to the most relevant case, in
which the number of files $N$ is larger than or equal to the number of
users $K$.  The main results are presented later for the general case.

In this setting, the rate, i.e.  the load of the shared link
normalized by the file size, in the delivery phase of conventional
uncoded caching schemes is
\begin{equation*}
    K\cdot(1-M/N).
\end{equation*}
Here $K$ is the rate without caching, and $1-M/N$ is the local caching
gain.  In contrast, the coded caching scheme proposed in this paper
attains a rate of
\begin{equation*}
    K\cdot(1-M/N)\cdot\frac{1}{1+KM/N}.
\end{equation*}
Thus, in addition to the local caching gain of $1-M/N$, coded caching
also achieves a global caching gain of $\tfrac{1}{1+KM/N}$. Both of
these gains indicate the multiplicative reduction in rate of the shared
link, so that a smaller factor means a larger rate reduction.  

Observe that the local caching gain depends on the normalized
\emph{local} cache size $M/N$ and is relevant only if the cache size $M$
is on the order of the number of files $N$. On the other hand, the
global caching gain depends on the normalized \emph{cumulative} cache
size $KM/N$ and is relevant whenever the cumulative cache size $KM$ is on
the order of (or larger than) the number of files $N$. 

By deriving fundamental lower bounds on the rate required in the
delivery phase, we show that the rate of the proposed coded caching
scheme is within a factor $12$ of the information-theoretic optimum for
all values of $N$, $K$, and $M$.

\begin{figure}[htbp]
    \centering
    \hspace{-1.8cm}\includegraphics{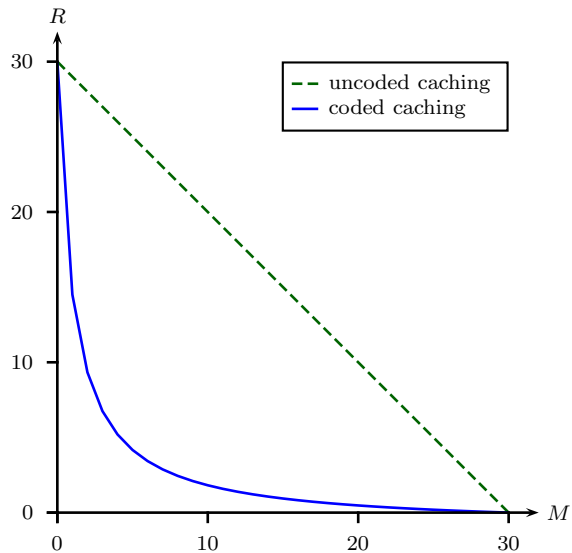}
    \caption{Rate $R$ required in the delivery phase as a function of
        memory size $M$ for $N=30$ files and $K=30$ users. The figure
        compares the performance of the proposed coded caching scheme 
        to that of conventional uncoded caching.}
    \label{fig:tradeoff}
\end{figure}

To obtain some intuition for the effect of these two gains, let us
compare the rates of the conventional uncoded scheme (achieving only the
local gain) versus the proposed coded scheme (achieving both the local
and global gains) for a system with $N=30$ files and $K=30$ users as
shown in Fig.~\ref{fig:tradeoff}. When each user has a cache memory
large enough to store $M=10$ files, the rate of the uncoded caching
scheme corresponds to sending $20$ files over the shared link, while the
proposed coded caching scheme achieves a rate corresponding to sending
only $1.8$ files: a reduction by a factor $11$ in rate.

The remainder of this paper is organized as follows.
Section~\ref{sec:problem} formally introduces our information-theoretic
formulation of the caching problem.  Section~\ref{sec:main} presents
main results, which are illustrated with examples in
Section~\ref{sec:examples}. Sections~\ref{sec:scheme}--\ref{sec:approx}
contain proofs. Section~\ref{sec:discussion} discusses some follow-up
results and directions for future research.

\section{Problem Setting}
\label{sec:problem}

Before formally introducing the problem in
Section~\ref{sec:problem_formal}, we start with an informal description
in Section~\ref{sec:problem_informal}.

\subsection{Informal Problem Description}
\label{sec:problem_informal}

We consider a system with one server connected through a shared,
error-free link to $K$ users, as shown in Fig.~\ref{fig:setting}. The
server has access to a database of $N$ files $W_1, \dots, W_N$ each of
size $F$ bits. Each user $k$ has an isolated cache memory $Z_k$ of size
$MF$ bits for some real number $M\in[0,N]$. 

The system operates in two phases: a \emph{placement phase} and a
\emph{delivery phase}. In the placement phase, the users are given
access to the entire database $W_1, \dots, W_N$ of files. Each user $k$
is then able to fill the content of its cache $Z_k$ using the database.
In the delivery phase, only the server has access to the database of
files. Each user $k$ requests one of the files $W_{d_k}$ in the
database. The server is informed of these requests and proceeds by
transmitting a signal $X_{(d_1, \ldots, d_k)}$ of size $RF$ bits over
the shared link for some fixed real number $R$. The quantities $RF$ and
$R$ are referred to as the load and the rate of the shared link,
respectively. Using the content $Z_k$ of its cache and the signal
$X_{(d_1,\ldots, d_k)}$ received over the shared link, each user $k$
aims to reconstruct its requested file $W_{d_k}$. 

A memory-rate pair $(M,R)$ is \emph{achievable for requests
$d_1,\dots,d_K$} if every user $k$ is able to recover its desired file
$W_{d_k}$ (with high probability for $F$ large enough). A memory-rate
pair $(M,R)$ is said to be \emph{achievable} if this pair is achievable
for every possible request $d_1,\dots,d_K$ in the delivery phase.
Finally, we denote by $R^\star(M)$ the smallest rate $R$ such that
$(M,R)$ is achievable. The function $R^\star(M)$ describes the
\emph{memory-rate tradeoff} for the caching problem. The aim of this
paper is to characterize this memory-rate tradeoff. In other words, we
aim to find the minimum rate of communication over the shared link at
which all possible demand tuples can be satisfied.

We illustrate these definitions with the example of the caching strategy
employed by conventional uncoded caching systems, which will serve as a
baseline scheme throughout the remainder of this paper.

\begin{example}[\emph{Uncoded  Caching}]
    \label{eg:problem}
    For a memory size of $MF$ bits, one possible strategy is for each
    user to cache the same $M/N$ fraction of each file in the placement
    phase.  In the delivery phase, the server simply transmits the
    remaining $1-M/N$ fraction of any requested file over the shared
    link.  Clearly, each user can recover its requested file from the
    content of its local cache and the signal sent over the shared link.
    In the worst case the users request different files---the delivery
    rate for this caching scheme is thus
    \begin{equation}
        \label{eq:uncoded}
        R_U(M) \defeq K \cdot(1-M/N) \cdot \min \{1, N/K\}  
    \end{equation}
    We refer to this caching strategy as \emph{uncoded caching}, since
    both content placement and delivery are uncoded. 

    The first factor $K$ in~\eqref{eq:uncoded} is the rate without
    caching. The second factor in~\eqref{eq:uncoded} is $1-M/N$. We call
    this the \emph{local caching gain}, since it arises from having a
    fraction $M/N$ of each file available locally at the user. If $N <
    K$, the system enjoys an additional gain, reflected in the third
    factor in~\eqref{eq:uncoded}. In this case, some users will
    necessarily request the same file, resulting in a natural
    multicasting gain of $N/K$. 
\end{example}

\subsection{Formal Problem Statement}
\label{sec:problem_formal}

We now provide the formal definition of the information-theoretic
caching problem. Let $(W_n)_{n=1}^N$ be $N$ independent random variables
each uniformly distributed over
\begin{equation*}
    [2^F] \defeq \{1, 2, \dots, 2^F\}
\end{equation*}
for some $F\in\N$. Each $W_n$ represents a file of size $F$ bits. A
$(M,R)$ caching scheme consists of $K$ caching functions, $N^K$ encoding
functions, and $K N^K$ decoding functions. 

The $K$ caching functions 
\begin{equation*}
    \phi_k\from [2^{F}]^{N} \to [2^{\floor{FM}}]
\end{equation*}
map the files $W_1, \dots, W_N$ into the cache content
\begin{equation*}
    Z_k \defeq \phi_k(W_1,\dots W_N)
\end{equation*}
for each user $k\in[K]$ during the placement phase. The $N^K$ encoding functions
\begin{align*}
    \psi_{(d_1,\dots, d_K)}\from [2^{F}]^{N} \to [2^{\floor{FR}}]
\end{align*}
map the files $W_1, \dots, W_N$ to the input
\begin{align*}
    X_{(d_1, \ldots, d_K)} \defeq \psi_{(d_1, \ldots, d_K)}(W_1, \ldots, W_N)
\end{align*}
of the shared link responding to the requests $(d_1, \ldots, d_K) \in [N]^K$ during
the delivery phase. Finally, the $K N^K$ decoding functions
\begin{align*}
    \mu_{(d_1,\ldots, d_K),k}\from [2^{\floor{RF}}] \times  [2^{\floor{FM}}] \to  [2^{F}]
\end{align*}
map the signal received over the shared link $X_{(d_1, \ldots, d_K)}$
and the cache content $Z_k$ to the estimate
\begin{equation*}
    \hat{W}_{(d_1,\ldots, d_K), k} 
    \defeq \mu_{(d_1,\ldots, d_K),k}(X_{(d_1, \ldots, d_K)}, Z_k)
\end{equation*}
of the requested file $W_{d_k}$ of user $k\in[K]$. The probability of
error is defined as
\begin{equation*}
    \max_{(d_1,\dots,d_K)\in[N]^K} \max_{k\in[K]}
    \Pp\big(
    \hat{W}_{(d_1,\ldots, d_K), k} \neq W_{d_k}
    \big).
\end{equation*}

\begin{definition}
    The pair $(M,R)$ is \emph{achievable} if for every $\varepsilon>0$
    and every large enough file size $F$ there exists a $(M,R)$ caching
    scheme with probability of error less than $\varepsilon$. We define
    the \emph{memory-rate tradeoff}
    \begin{equation*}
        R^\star(M) \defeq \inf \big\{R: (M,R) \text{ is achievable} \big\}.
    \end{equation*}
\end{definition}

\section{Main Results}
\label{sec:main}

The first theorem presents an achievable rate $R_C(M)$, yielding an
upper bound on the memory-rate tradeoff $R^\star(M)$.

\begin{theorem}
    \label{thm:achievability}
    For $N\in\N$ files and $K \in\N $ users each with cache of size 
    $M\in\{0, N/K, 2N/K, \dots, N\}$, 
    \begin{equation*}
        R^\star(M) 
        \leq R_C(M)
        \defeq K\cdot(1-M/N)\cdot\min\Big\{\frac{1}{1+KM/N},\frac{N}{K}\Big\}
    \end{equation*}
    is achievable. For general $0 \leq M \leq N$, the lower convex
    envelope of these points is achievable.
\end{theorem}

The rate $R_C(M)$ is achieved by a coded caching scheme that is
described and analyzed in detail in Section~\ref{sec:scheme}. For ease
of exposition, we first focus on the case $N \geq K$, in which
\begin{equation} 
    \label{eq:achievabilityN>K}
    R_C(M)
    = K\cdot(1-M/N)\cdot \frac{1}{1+KM/N}.
\end{equation}
The achievable rate $R_C(M)$ consists of three distinct factors.
The first factor in $R_C(M)$ is $K$.  This is the worst-case rate without
caches at the users (i.e., $M=0$).

The second factor in $R_C(M)$ is $1-M/N$. Referring
to~\eqref{eq:uncoded} in Example~\ref{eg:problem}, we see that this
term, capturing the local caching gain, appears also in the rate
expression of the uncoded caching scheme. Observe that this local gain
is a function of the normalized \emph{local} memory size $M/N$, and it
is relevant whenever $M$ is on the order of $N$. 

Finally, the third factor in $R_C(M)$ is $\tfrac{1}{1+KM/N}$, which we
call the \emph{global caching gain}.  This gain is a function of the
normalized \emph{global} or \emph{cumulative} memory size $KM/N$, and it
is relevant whenever $KM$ is on the order of (or larger than) $N$. This
global gain is to be interpreted as a multicasting gain available
simultaneously for all possible demands. Note that, since the number of
users is smaller than the number of files, in the worst case all users
request different files. Hence, there are no natural multicasting
opportunities. The scheme proposed in Theorem~\ref{thm:achievability}
carefully designs the content placement in order to create coded
multicasting opportunities in the delivery phase even among users that
request different files. Since the placement phase is performed without
knowledge of the actual demands, care must be taken to ensure that the
same multicasting opportunities are created simultaneously for every
possible set of requests in the delivery phase.

We point out that the uncoded caching scheme introduced in
Example~\ref{eg:problem} achieves only the local caching gain, whereas
the coded caching scheme proposed in Theorem~\ref{thm:achievability}
achieves both the local as well as the global caching gains.  The
following two examples compare these two gains.

\begin{example}[\emph{$\Theta(K)$ Improvement in Rate}] 
    \label{eg:slpe/k} 
    Consider a system with the same number of users as files, i.e.,
    $N=K$. Assume each user has enough cache memory for half of the
    files so that $M = N/2$. Then the local caching gain is $1/2$ and
    the global caching gain is $1/(1+K/2)$. By \eqref{eq:uncoded},
    uncoded caching achieves a rate of $K/2$. On the other hand, by
    Theorem~\ref{thm:achievability}, coded caching achieves a rate of
    $(K/2)/(1+K/2)<1$: a reduction by more than a factor $K/2$ in rate
    compared to the uncoded scheme. We refer the reader to
    Fig.~\ref{fig:tradeoff} in Section~\ref{sec:introduction} for a
    visualization of the effect of this improvement.
\end{example}

\begin{example}[\emph{$\Theta(K)$ Improvement in Slope}] 
    \label{eg:rate/k} 
    In this example, we compare the performance of the coded and uncoded
    caching schemes for small values of the cache size $M$.  We consider
    again the case $N=K$. From~\eqref{eq:uncoded}, the rate of uncoded
    caching has a slope of $-1$ around $M=0$.  On the other hand, by
    Theorem~\ref{thm:achievability}, the rate of coded caching has a
    slope less than $-K/2$ around $M=0$.\footnote{This follows by
        calculating the slope of the straight line connecting the two
    consecutive corner points of $R_C(M)$ at $M=0$ and $M=1$.}
    Therefore, the coded caching scheme reduces the rate over the shared
    link at least $K/2$ times faster as a function of cache size than
    the uncoded caching scheme. Comparing the rates of the uncoded and
    coded schemes in Fig.~\ref{fig:tradeoff} in
    Section~\ref{sec:introduction} for small values of $M$ illustrates
    the effect of this improvement.  
\end{example}

Consider next the case $N < K$, in which the third factor in $R_C(M)$ is
the minimum of $\tfrac{1}{1+KM/N}$ and $N/K$. The first term in this
minimum is the coded multicasting gain created by careful content
placement as discussed for the case $N \geq K$.  However, for a scenario
with fewer files than users, there exists already a natural multicasting
opportunity: by multicasting all $N$ files to the $K$ users, we can
achieve a gain of $N/K$. This is the second term in the minimum above.
The scheme in Theorem~\ref{thm:achievability} achieves the better of
these two gains. We point out that for $M \geq 1-N/K$ this minimum is
achieved by the first of the two gains. In other words, the natural
multicasting gain is relevant only when the memory size is very small.

Having established an upper bound on the memory-rate tradeoff
$R^\star(M)$, we proceed with a lower bound on it.

\begin{theorem}
    \label{thm:converse}
    For $N\in\N$ files and $K\in\N$ users each with cache of size $0 \leq M \leq
    N$,
    \begin{equation*}
        R^\star(M) \geq \max_{s\in\{1, \dots, \min\{N,K\}\}}
        \Bigl( s - \frac{s}{\floor{N/s}}M \Bigr).
    \end{equation*}
\end{theorem}

The proof of Theorem~\ref{thm:converse}, presented in
Section~\ref{sec:converse}, is based on a cut-set bound argument.
Tighter lower bounds on $R^\star(M)$ can be derived using stronger
arguments than the cut-set bound (see the discussion in
Example~\ref{eg:2x2} in Section~\ref{sec:examples}). However, the
cut-set bound alone is sufficient for a constant-factor approximation of
the memory-rate tradeoff $R^\star(M)$, as the next theorem shows by
comparing the achievable rate $R_C(M)$ in
Theorem~\ref{thm:achievability} with the lower bound in
Theorem~\ref{thm:converse}.

\begin{theorem}
    \label{thm:approx}
    For $N\in\N$ files and $K\in\N$ users each with cache of size $0 \leq M \leq
    N$,
    \begin{equation*}
        1 \leq \frac{R_C(M)}{R^\star(M)} \leq 12,
    \end{equation*}
    with the achievable rate $R_C(M)$ of coded caching as defined in
    Theorem~\ref{thm:achievability}.
\end{theorem}

The proof of Theorem~\ref{thm:approx} is presented in
Section~\ref{sec:approx}. The bound $R_C(M)/R^\star(M)\leq
12$ on the approximation ratio of the proposed caching scheme is
somewhat loose due to the analytical bounding techniques that were used.
Numerical simulations suggest that
\begin{equation*}
    \frac{R_C(M)}{R^\star(M)} \leq 5
\end{equation*}
for all $N$, $K$, and $0 \leq M \leq N$.

Theorem~\ref{thm:approx} shows that the rate $R_C(M)$ of the proposed
coded caching scheme in Theorem~\ref{thm:achievability} is close to the
information-theoretic optimum $R^\star(M)$ for all values of the system
parameters. More precisely, it shows that no scheme can improve upon the
rate $R_C(M)$ of the proposed scheme by more than a factor $12$.  This
also suggests that the local and global caching gains identified in this
paper are fundamental: there are no other significant caching gains
(i.e., scaling with the problem parameters) beyond these two.

\section{Examples}
\label{sec:examples}

\begin{example}
    \label{eg:2x2} 
    \begin{figure}[htbp]
        \centering 
        \includegraphics{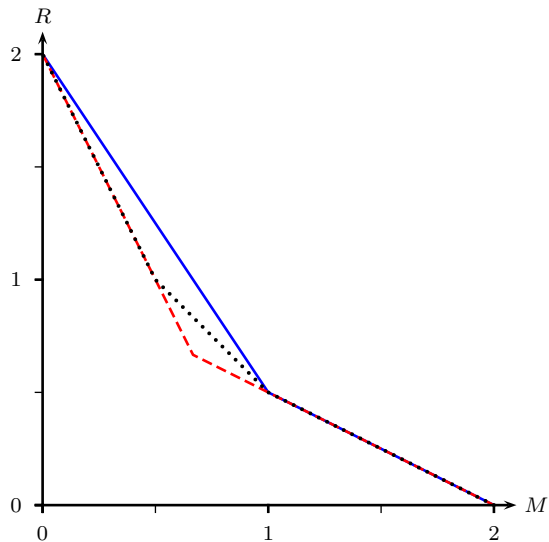} 
        \caption{Memory-rate tradeoff for $N=2$ files $K=2$ users. The
            achievable rate $R_C(M)$ of the coded caching scheme from
            Theorem~\ref{thm:achievability} is indicated by the solid
            blue curve. The lower bound on $R^\star(M)$ from
            Theorem~\ref{thm:converse} is indicated by the dashed red
            curve.  For the $N=K=2$ case, $R^\star(M)$ can be found exactly
            and is indicated by the dotted black curve.} 
        \label{fig:2x2}
    \end{figure}
    Consider the case $N=K=2$, so that there are two files, say $W_1=A,
    W_2=B$, and two users each with cache memory of size $M$. The upper
    and lower bounds in Theorems~\ref{thm:achievability}
    and~\ref{thm:converse} on the memory-rate tradeoff $R^\star(M)$ are
    depicted in Fig.~\ref{fig:2x2}. To illustrate the proof techniques,
    we now show how these two bounds are derived for this simple
    setting.

    We start with the upper bound in Theorem~\ref{thm:achievability},
    focusing on the corner points of the achievable region.   First, let
    us consider the two extreme cases $M=0$ and $M=N$.  If $M=0$, the
    server can always transmit both files $A$ and $B$ over the shared
    link. Since this satisfies every possible request, the $(M,R)$ pair
    $(0,2)$ is achievable.  If $M=2$, each user can cache both files $A$
    and $B$ in the placement phase. Therefore, no communication is
    needed in the delivery phase and the $(M,R)$ pair $(2,0)$ is
    achievable.

    \begin{figure}[htbp]
        \centering
        \hspace{0.1cm}\includegraphics{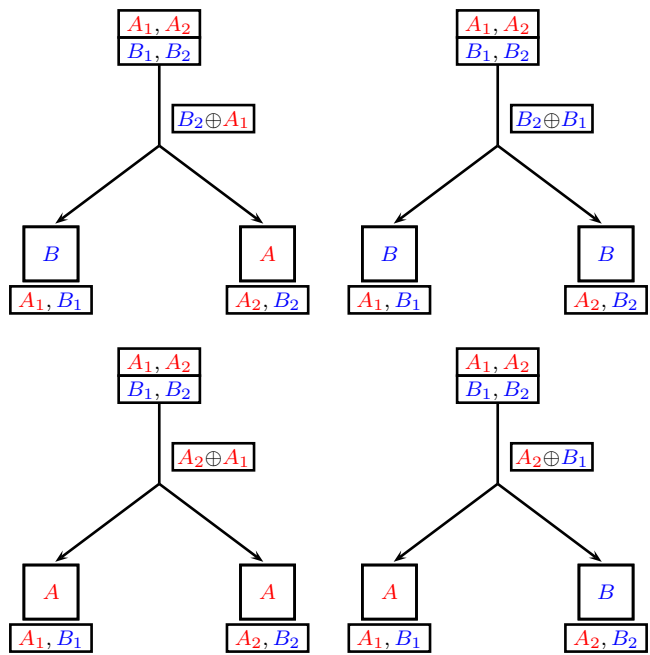}
        \caption{Caching strategy for $N=2$ files and $K=2$ users with
            cache size $M=1$ with all four possible user requests. Each
            file is split into two subfiles of size $1/2$, i.e.,
            $A=(A_1,A_2)$ and $B=(B_1,B_2)$. The scheme achieves rate
            $R=1/2$. Observe that, while the transmission from the
            server changes as a function of the user requests, the cache
            contents do not.} 
        \label{fig:solution_2x2}
    \end{figure}
    Consider then the more interesting corner point at $M=1$.
    The caching scheme achieving the upper bound in
    Theorem~\ref{thm:achievability} is as follows (see
    Fig.~\ref{fig:solution_2x2}). We split both files $A$ and $B$ into
    two subfiles of equal size, i.e., $A=(A_1, A_2)$ and $B=(B_1, B_2)$.
    In the placement phase, we set $Z_1=(A_1, B_1)$ and $Z_2=(A_2,B_2)$.
    In words, each user caches one exclusive part of each file. For the
    delivery phase, assume for example that user one requests file $A$
    and user two requests file $B$. Given that user one already has
    subfile $A_1$ of $A$, it only needs to obtain the missing subfile
    $A_2$, which is cached in the second user's memory $Z_2$. Similarly,
    user two only needs to obtain the missing subfile $B_1$, which is
    cached in the first user's memory $Z_1$. In other words, each user
    has one part of the file that the other user needs.

    The server can in this case simply transmit $A_2\oplus B_1$, where
    $\oplus$ denotes bitwise XOR.  Since user one already has $B_1$, it
    can recover $A_2$ from $A_2\oplus B_1$. Similarly, since user two
    already has $A_2$, it can recover $B_1$ from $A_2\oplus B_1$. Thus,
    the signal $A_2\oplus B_1$ received over the shared link helps both
    users to effectively exchange the missing subfiles available in the
    cache of the other user. 

    The signals sent over the shared link for all other requests are
    depicted in Fig.~\ref{fig:solution_2x2}. One can see that in all
    cases the signal is constructed using the same logic of exchanging
    the missing subfiles. This proves the achievability of the
    $(M,R)$ pair $(1,1/2)$.

    It is worth pointing out that in each case the server sends a
    single coded multicast transmission to satisfy two (possibly
    different) user requests. Moreover, these coded multicasting
    opportunities are available simultaneously for all four possible
    user requests. This availability of simultaneous multicasting
    opportunities, enabled by careful content placement, is critical,
    since the placement phase has to be performed without knowledge of
    the actual demands in the delivery phase. 

    So far, we have shown that the $(M,R)$ pairs at corner points
    $(0,2)$, $(1,1/2)$, and $(2, 0)$ are achievable.  On the other hand,
    by dividing the cache memories and the transmitted signal
    proportionally, it is easy to see that if any two points $(M_1,
    R_1)$ and $(M_2, R_2)$ are achievable, then the line connecting them
    is also achievable. Inspired by the term time sharing in network
    information theory, we refer to this as \emph{memory sharing}.
    
    Memory sharing between the corner points $(0,2)$, $(1,1/2)$, and
    $(2, 0)$ establishes the achievability of the solid blue curve in
    Fig.~\ref{fig:2x2}, which coincides with the upper bound stated in
    Theorem~\ref{thm:achievability}.

    We continue by analyzing the lower bound on $R^\star(M)$ in
    Theorem~\ref{thm:converse}. The proof relies on the cut-set bound.
    We consider two cuts. The first cut separates $(X_{(1,2)}, Z_1,
    Z_2)$ from the two users. Assume $(M,R)$ is an achievable
    memory-rate pair. Then this cut has capacity at most $RF+2MF$, since
    $X_{(1,2)}$ is at most $RF$ bits by definition of achievability, and
    since $Z_1, Z_2$ contain each $MF$ bits. On the other hand, since
    the first user can recover $A$ from $(X_{(1,2)},Z_1)$ and the second
    user can recover $B$ from $(X_{(1,2)},Z_2)$, the number of bits that
    need to be transmitted over this cut is at least $2F$. Hence,
    \begin{align*}
        RF + 2MF \geq 2F,
    \end{align*}
    so that
    \begin{align*}
        R \geq 2-2M.
    \end{align*}
    As this holds for all achievable memory-rate pairs $(M,R)$, we
    conclude that
    \begin{equation*}
        R^\star(M) \geq 2-2M.
    \end{equation*}

    The second cut separates $(X_{(1,2)}, X_{(2,1)}, Z_1)$ from the
    first user. Note that this user can recover $A$ and $B$ from
    $(X_{(1,2)}, Z_1)$ and $(X_{(2,1)}, Z_1)$, respectively. Hence, this
    cut yields
    \begin{align*}
        2RF + MF \geq 2F
    \end{align*}
    for any achievable memory-rate pair $(M,R)$, implying that
    \begin{equation*}
        R^\star(M) \geq 1-M/2.
    \end{equation*}
    Together, this yields the dashed red curve in Fig.~\ref{fig:2x2},
    which coincides with the lower bound stated in
    Theorem~\ref{thm:converse}.

    For the case $N=K=2$, the memory-rate tradeoff can in fact be found
    exactly and is indicated by the dotted black curve in
    Fig.~\ref{fig:2x2}.  This is argued by showing that the pair
    $(M,R)=(1/2,1)$ is also achievable, and by deriving the additional
    non cut-set bound
    \begin{equation*}
        R^\star(M) \geq 3/2-M.
    \end{equation*}
    This shows that, while the bounds in
    Theorems~\ref{thm:achievability} and~\ref{thm:converse} are
    sufficient to characterize the memory-rate tradeoff $R^\star(M)$ to
    within a constant multiplicative gap, neither the achievable region
    nor the lower bound are tight in general. The details of this
    derivation are reported in the Appendix.
\end{example}

\begin{example}
    \label{eg:3x3}
    \begin{figure}[htbp]
        \centering 
        \includegraphics{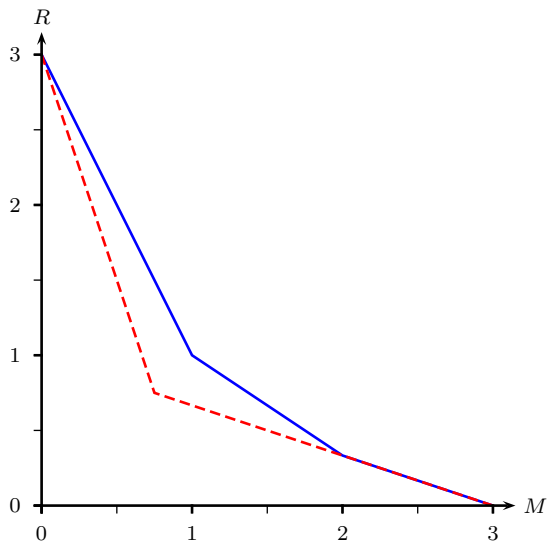} 
        \caption{Memory-rate tradeoff for $N=3$ files and $K=3$ users.
        The achievable rate $R_C(M)$ of the coded caching scheme from 
        Theorem~\ref{thm:achievability} is indicated by the
        solid blue curve. The lower bound on $R^\star(M)$ from
        Theorem~\ref{thm:converse} is indicated by the dashed red
        curve.} 
        \label{fig:3x3}
    \end{figure}
    In this example, we assume that $N=K=3$ so that there are three
    users and three files, say $W_1=A$, $W_2=B$, and $W_3=C$. Again, it
    is trivial to see that the  $(M,R)$ pairs $(0,3)$ and $(3,0)$
    are achievable. We focus on two nontrivial corner points at $M=1$ and $M=2$.

    Consider first caches of size $M=1$. We split each file into three
    subfiles of equal size, i.e., $A=(A_1, A_2, A_3)$, $B=(B_1, B_2,
    B_3)$, and $C=(C_1, C_2, C_3)$. In the placement phase, the cache
    content of user $k$ is selected as $Z_k = (A_k, B_k, C_k)$. A more
    formal way to describe this content placement, which is a bit
    exaggerated for this simple setting but will be useful for the
    general setting below, is as follows. Let $\mc{T}$ be a subset of
    one element of $\{1,2,3\}$. Then subfiles $A_{\mc{T}}$,
    $B_{\mc{T}}$, $C_{\mc{T}}$ are placed into the cache of user $k$ if
    $k\in\mc{T}$. For example, $A_{1}$ is cached at user one since in
    this case $\mc{T} = \{1\}$. 

    For the delivery phase, let us consider as an example that user one
    requests file $A$, user two requests file $B$, and user three
    requests file $C$. Then the missing subfiles are $A_2$ and $A_3$ for
    user one, $B_1$ and $B_3$ for user two, and $C_1$ and $C_2$ for user
    three.  Given the cache contents, users one and two aim to exchange
    $A_2$ and $B_1$, users one and three aim to exchange $A_3$ and
    $C_1$, and users two and three aim to exchange $B_3$ and $C_2$.  The
    signal $(A_2 \oplus B_1, A_3 \oplus C_1, B_3 \oplus C_2)$ enables
    all of these three exchanges.  All other requests can be satisfied
    in a similar manner. Since each subfile has rate $1/3$, the proposed
    scheme achieves a rate of $1$, and therefore $(M,R)$ pair $(1,1)$ is
    achievable.

    Observe that, through careful content placement, we have again
    created coded multicasting opportunities for any two users even with
    different demands. Moreover, these coded multicasting opportunities
    are available simultaneously for all $27$ possible triples of user
    requests.

    Consider next caches of size $M=2$. We  again split each file into
    three subfiles of equal size. However, it will be convenient to
    label these subfiles differently, namely $A=(A_{12}, A_{13},
    A_{23})$, $B=(B_{12}, B_{13}, B_{23})$, and $C=(C_{12}, C_{13},
    C_{23})$. In the placement phase, the caching strategy is
    \begin{align*}
        Z_1 & = (A_{12}, A_{13}, B_{12}, B_{13}, C_{12}, C_{13}),\\
        Z_2 & = (A_{12}, A_{23}, B_{12}, B_{23}, C_{12}, C_{23}),\\
        Z_3 & = (A_{13}, A_{23}, B_{13}, B_{23}, C_{13}, C_{23}).
    \end{align*}
    This content placement can be understood as follows. Let $\mc{T}$ be a
    subset of two elements of $\{1,2,3\}$. Then subfiles $A_{\mc{T}}$,
    $B_{\mc{T}}$, $C_{\mc{T}}$ are placed into the cache of user $k$ if
    $k\in\mc{T}$. For example, $A_{13}$ is cached at users one and three
    since in this case $\mc{T} = \{1,3\}$. 

    For the delivery phase, let us again assume as an example that user
    one requests file $A$, user two requests file $B$, and user three
    requests file $C$. In this case, user one misses subfile $A_{23}$,
    which is available at both users two and three. User two misses
    subfile $B_{13}$, which is available at both users one and three.
    And user three misses subfile $C_{12}$, which is available at both
    users one and two. In other words, the three users would like to
    exchange the subfiles $A_{23}, B_{13}, C_{12}$. This exchange can be
    enabled by transmitting the signal $A_{23} \oplus B_{13} \oplus
    C_{12}$ over the shared link. Given its cache content, each user can
    then recover the missing subfile. All other requests can be
    satisfied in a similar manner. The rate of transmission in the
    delivery phase is $1/3$, and therefore $(M,R)$ pair $(2,1/3)$ is
    achievable. This approach again creates simultaneous coded
    multicasting opportunities, but this time for all three users
    together.

    The arguments so far show that the solid blue curve in
    Fig.~\ref{fig:3x3} is achievable. This coincides with the upper bound
    in Theorem~\ref{thm:achievability}.

    For the lower bound on $R^\star(M)$, we use two cut-set bounds.  The
    first cut separates $(X_{(1,2,3)}, Z_1, Z_2, Z_3)$ from the three
    users. Note that the users can recover $A$, $B$, and $C$, from
    $(X_{(1,2,3)},Z_1)$, $(X_{(1,2,3)},Z_2)$, and $(X_{(1,2,3)},Z_3)$,
    respectively. For any achievable $(M,R)$ pair, the capacity of this
    cut is at most $RF+3MF$, and the number of bits that need to be
    transmitted over it is at least $3F$. Hence,
    \begin{align*}
        RF+3MF \geq 3F,
    \end{align*} 
    which implies that
    \begin{equation*}
        R^\star(M) \geq 3-3M.
    \end{equation*}

    The second cut separates $(X_{(1,2,3)}, X_{(2,3,1)}, X_{(3,1,2)},
    Z_1)$ from the first user. Note that this user can recover $A$, $B$,
    and $C$ from $(X_{(1,2,3)}, Z_1)$, $(X_{(2,3,1)}, Z_1)$, and
    $(X_{(3,1,2)}, Z_1)$, respectively. For any achievable $(M,R)$
    pair, the cut capacity is at most $3RF+MF$, and the number of bits
    that need to be transmitted over it is at least $3F$. Hence,
    \begin{align*} 
        3RF+MF \geq 3F,
    \end{align*} 
    which implies that
    \begin{equation*}
        R^\star(M) \geq 1-M/3.
    \end{equation*}

    Together, these two cut-set bounds result in the dashed red curve in
    Fig.~\ref{fig:3x3}. This coincides with the lower bound in
    Theorem~\ref{thm:converse}.
\end{example}

\section{Coded Caching Scheme (Proof of Theorem~\ref{thm:achievability})}
\label{sec:scheme}

We now present the general achievable scheme. We first describe the
algorithm in words, focusing on the corner points of $R_C(M)$. Consider
cache size $M\in\{0, N/K, 2N/K, \dots, N\}$, and set
\begin{equation*}
    t \defeq MK/N.
\end{equation*}
Observe that $t$ is an integer between $0$ and $K$. 

If $M=0$, then in the delivery phase the server can simply transmit the
union of all requested files over the shared link, resulting in
$F\min\{N,K\}$ bits being sent. Hence 
\begin{equation*}
    R^\star(0) \leq \min\{N,K\}. 
\end{equation*}
If $M=N$, then all files in the database can be cached at every user in
the placement phase. Hence 
\begin{equation*}
    R^\star(N) = 0.
\end{equation*}
Assume in the following that $M$ is strictly larger than zero and
strictly less than $N$, so that $t\in\{1,2,\dots,K-1\}$. 

In the placement phase, each file is split into $\tbinom{K}{t}$
nonoverlapping subfiles
of equal size. It will be convenient to label the subfiles of file $W_n$
as
\begin{equation*}
    W_n = (W_{n,\mc{T}}: \mc{T} \subset [K], \card{\mc{T}}=t),
\end{equation*}
where we recall the notation 
\begin{align*}
    [K] & \defeq \{1,\dots, K\}.
\end{align*}
For each $n\in[N]$, subfile $W_{n,\mc{T}}$ is placed in the cache of
user $k$ if $k\in\mc{T}$. Thus, each user caches a total of
$N\tbinom{K-1}{t-1}$ subfiles. Since each of these subfiles has size
$F/\tbinom{K}{t}$, this requires
\begin{equation*}
    N\tbinom{K-1}{t-1}\frac{F}{\tbinom{K}{t}} =  F\frac{Nt}{K} = FM
\end{equation*}
bits of cache memory at each user, satisfying the memory constraint.

\begin{example}
    \label{eg:3x3b}
    Let $N=K=3$ and $M=2$. Then $t=2$ and the content placement is
    \begin{align*}
        Z_1 & = (W_{n,\{1,2\}}, W_{n,\{1,3\}})_{n=1}^3,\\
        Z_2 & = (W_{n,\{1,2\}}, W_{n,\{2,3\}})_{n=1}^3,\\
        Z_3 & = (W_{n,\{1,3\}}, W_{n,\{2,3\}})_{n=1}^3,
    \end{align*}
    as we have already seen in Example~\ref{eg:3x3} in
    Section~\ref{sec:examples}.
\end{example}

We next describe the delivery phase. Consider the request vector
$(d_1,\dots, d_K)\in [N]^K$, i.e., user $k$ requests file $W_{d_k}$.  We
focus on a subset $\mc{S}\subset [K]$ of $\card{\mc{S}}=t+1$ users.
Observe that every $t$ users in $\mc{S}$ share a subfile in their caches
that is needed at the remaining user in $\mc{S}$. More precisely, fix a
user $s \in \mc{S}$, and note that $\card{\mc{S}\setminus\{s\}}=t$. The
subfile $W_{d_s,\mc{S}\setminus\{s\}}$ is requested by user $s$ since it
is a subfile of $W_{d_s}$. At the same time, it is missing at user $s$
since $s\notin\mc{S}\setminus\{s\}$.  Finally, it is present in the
cache of any user $k\in\mc{S}\setminus\{s\}$. 

For each such subset $\mc{S}\subset [K]$ of cardinality $\card{\mc{S}}=t+1$, 
the server transmits
\begin{equation*}
    \oplus_{s \in\mc{S}} W_{d_{s},S\setminus\{s\}},
\end{equation*}
where $\oplus$ denotes bitwise XOR.  Each of these
sums results in $F/\tbinom{K}{t}$ bits being sent over the shared link.
Since the number of subsets $\mc{S}$ is $\tbinom{K}{t+1}$, the total
number of bits sent over the shared link is
\begin{align*}
    FR 
    & = \tbinom{K}{t+1}\frac{F}{\tbinom{K}{t}} \\
    & = F\frac{K-t}{t+1} \\
    & = F\frac{K(1-M/N)}{1+KM/N},
\end{align*}
where we have used that $t=MK/N$.

\begin{example}
    \label{eg:3x3c}
    Let $N=K=3$ and $M=1$. Then $t=1$ and the content placement is $Z_k =
    (W_{n,\{k\}})_{n=1}^3$. For request $(d_1,d_2,d_3)=(1,2,3)$, the
    signal transmitted in the delivery phase is
    \begin{align*}
        X_{(1,2,3)} 
        = \big( 
        W_{1,\{2\}}  \oplus W_{2,\{1\}},\ 
        W_{1,\{3\}}\oplus W_{3,\{1\}},\ 
        W_{2,\{3\}}\oplus W_{3,\{2\}}
        \big),
    \end{align*}
    as we have already seen in Example~\ref{eg:3x3} in
    Section~\ref{sec:examples}.  Here, the three sums correspond to
    $\mc{S}=\{1,2\}$, $\mc{S}=\{3,1\}$, and $\mc{S}=\{3,2\}$,
    respectively. 
\end{example}

We now argue that each user can successfully recover its requested
message. Consider again a subset $\mc{S}\subset [K]$ with
$\card{\mc{S}}=t+1$. Since each user $k \in \mc{S}$ already has access
to the subfiles $W_{d_{s},S\setminus\{s\}}$ for all
$s \in \mc{S}\setminus\{k\}$, it can solve for
$W_{d_k,S\setminus\{k\}}$ from the signal
\begin{equation*}
    \oplus_{s\in\mc{S}} W_{d_{s},S\setminus\{s\}}
\end{equation*}
sent over the shared link. Since this is true for every such subset
$\mc{S}$, each receiver $k$ is able to recover all subfiles of the form
\begin{equation*}
    \big\{ W_{d_k,\mc{T}}: \mc{T}\subset [K]\setminus\{k\},
    \card{\mc{T}} = t
    \big\}
\end{equation*}
of the requested file $W_{d_k}$. The remaining subfiles are of the form
$W_{d_k,\mc{T}}$ for $\mc{T}$ such that $k\in\mc{T}$. But these subfiles
are already available in the cache of user $k$. Hence each user $k$ can
recover all subfiles of its requested file $W_{d_k}$. 

This shows that, for $M\in\{N/K, 2N/K, \dots, (K-1)N/K\}$, 
\begin{equation*}
    R^\star(M) \leq K\cdot(1-M/N)\cdot\frac{1}{1+KM/N}.
\end{equation*}
As was pointed out earlier (see Example~\ref{eg:2x2} in
Section~\ref{sec:examples}), the points on the line connecting any two
achievable points are also achievable. Finally, taking the minimum
between the rate derived here and the rate
\begin{equation*}
    K\cdot(1-M/N)\cdot\min\{1,N/K\}
\end{equation*}
achieved by the conventional uncoded scheme described in
Example~\ref{eg:problem} in Section~\ref{sec:problem_informal} proves
Theorem~\ref{thm:achievability}.\hfill\IEEEQED

For completeness, in Algorithm~\ref{alg:coded-caching}, we formally
describe the placement and the delivery procedures of the coded caching
scheme for $N$ files, $K$ users. We focus on the corner points of
$R_C(M)$, which occur at cache size $M$ such that $MK/N$ is a positive
integer less than $K$. 

\begin{algorithm}[h!]
    \caption{Coded Caching}
    \label{alg:coded-caching}
\begin{algorithmic}[0]
    \Statex
    \Procedure{Placement}{$W_1, \dots, W_N$}
    \State $t \gets MK/N$ 
    \State $\mathfrak{T} \gets \{\mc{T}\subset [K]: \card{\mc{T}} = t\}$
    \For{$n\in[N]$}
    \State split $W_n$ into $(W_{n,\mc{T}}: \mc{T}\in\mathfrak{T})$ of equal size
    \EndFor
    \For{$k\in[K]$}
    \State $Z_k \gets (W_{n,\mc{T}}: n\in[N], \mc{T}\in\mathfrak{T}, k\in\mc{T})$
    \EndFor
    \EndProcedure
    \Statex
    \Procedure{Delivery}{$W_1, \dots, W_N, d_1, \dots, d_K$}
    \State $t \gets MK/N$
    \State $\mathfrak{S} \gets \{\mc{S}\subset [K]: \card{\mc{S}} = t+1\}$
    \State $X_{(d_1,\dots,d_K)} \gets ({\textstyle\oplus_{k\in\mc{S}}} W_{d_k,S\setminus\{k\}}: \mc{S}\in\mathfrak{S})$
    \EndProcedure
\end{algorithmic}
\end{algorithm}

Observe that the placement phase of the algorithm is designed such that,
in the delivery phase, the caches enable coded multicasting between
$MK/N+1$ users with different demands. These coded multicasting
opportunities are available simultaneously for all $N^K$ possible user
demands. This is again critical, since the placement phase has to be
designed without knowledge of the actual demands in the delivery phase.

We also point out that while the problem setting allows for vanishing
probability of error as $F\to\infty$, the achievable scheme presented
here actually has zero error probability for finite $F$.

\section{Lower Bound on $R^\star(M)$ (Proof of Theorem~\ref{thm:converse})}
\label{sec:converse}

\begin{figure}[htbp]
    \centering
    \includegraphics{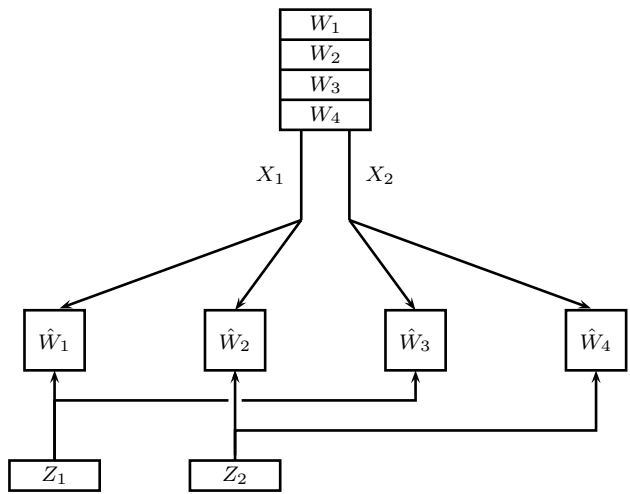}
    \caption{Cut corresponding to parameter $s=2$ in the proof of the
    converse. In the figure, $N=K=4$, and $X_1 = X_{(1,2,3,4)}$,
    $X_2=X_{(3,4,2,1)}$.} 
    \label{fig:converse}
\end{figure}

Let $s\in\{1,\dots, \min\{N,K\}\}$ and consider the first $s$ caches
$Z_1,\dots, Z_s$.  There exists a user demand and a corresponding input
to the shared link, say $X_1$, such that $X_1$ and $Z_1,\dots, Z_s$
determine the files $W_1,\dots, W_s$. Similarly, there exists an input
to the shared link, say $X_2$, such that $X_2$ and $Z_1,\dots, Z_s$
determine the files $W_{s+1},\dots, W_{2s}$. Continue in the same manner
selecting appropriate $X_3, \dots, X_{\floor{N/s}}$, see
Fig.~\ref{fig:converse}. We then have that $X_1,\dots, X_{\floor{N/s}}$
and $Z_1,\dots, Z_s$ determine $W_1,\dots, W_{s\floor{N/s}}$. 

Consider then the cut separating $X_1,\dots, X_{\floor{N/s}}$ and
$Z_1,\dots, Z_s$ from the corresponding users as indicated in
Fig.~\ref{fig:converse}. By the cut-set bound
\cite[Theorem~14.10.1]{cover91}, 
\begin{align*}
    \floor{N/s} R^\star(M)+sM \geq s\floor{N/s}.
\end{align*}
Solving for $R^\star$ and optimizing over all possible choices of $s$,
we obtain 
\begin{align*}
    R^\star(M) \geq \max_{s\in\{1,\dots, \min\{N,K\}\}}
    \Bigl( s - \frac{s}{\floor{N/s}}M \Bigr),
\end{align*}
proving the theorem. \hfill\IEEEQED

\section{Approximation of $R^\star(M)$ (Proof of Theorem~\ref{thm:approx})}
\label{sec:approx}

\begin{figure}[htbp]
    \centering 
    \hspace{-0.3cm}\includegraphics{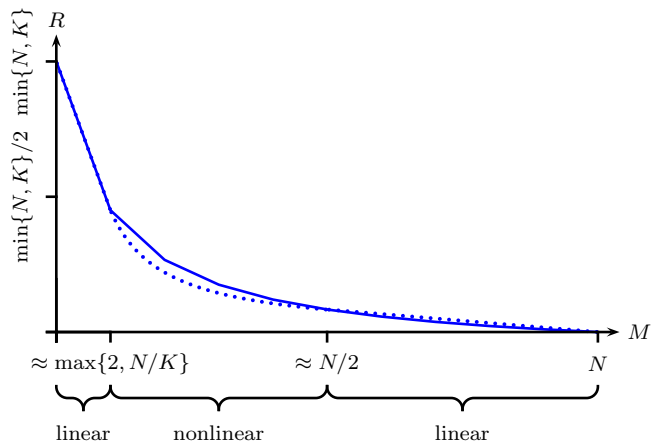} 
    \caption{Achievable rate $R_C(M)$
    (solid curve) as defined in Theorem~\ref{thm:achievability} for
    $N=20$ files and $K=10$ users.
    The function $R_C(M)$ has three distinct regimes, which are
    approximately $0 \leq M \leq \max\{2,N/K\}$, $\max\{2,N/K\} < M \leq
    N/2$, and $N/2 < M \leq N$. In the first and third regimes, $R_C(M)$
    is essentially linear; in the second regime, $R_C(M)$ is nonlinear (as
    indicated by the dotted curve).}
    \label{fig:regimes}
\end{figure}

We now compare the upper bound on $R^\star(M)$ in
Theorem~\ref{thm:converse} to the rate $R_C(M)$ achieved by the proposed
scheme given by the lower convex envelope of the points 
\begin{equation*}
    R_C(M) = \min\Big\{
    \frac{K(1-M/N)}{1+KM/N},N-M
    \Big\}
\end{equation*}
for $M$ a multiple of $N/K$ as described in Theorem~\ref{thm:achievability}.
The proof is based on the following observation. The function $R_C(M)$ has
three distinct regimes as depicted in Fig.~\ref{fig:regimes}. The first
regime is for $M$ between $0$ and approximately $\max\{2,N/K\}$. In this
regime, $R_C(M)$ is close to linear.  The second regime is for $M$
between approximately $\max\{2,N/K\}$ and $N/2$, in which $R_C(M)$ is
nonlinear and behaves essentially like $N/M$. The third regime is for
$M$ between approximately $N/2$ and $N$, in which $R_C(M)$ is again close
to linear.  We bound the ratio $R^\star(M)/R_C(M)$ separately in each of
these regimes---though, to optimize the constants, we will choose
slightly different definitions of the boundaries for the three regions. 

It will be convenient to consider the two cases $\min\{N,K\} \leq 12$
and $\min\{N,K\} \geq 13$ separately. We have
\begin{equation*}
    R_C(M) \leq \min\{N,K\}(1-M/N)
\end{equation*}
by Theorem~\ref{thm:achievability}. On the other hand, setting $s=1$ in
Theorem~\ref{thm:converse} yields
\begin{equation*}
    R^\star(M) \geq 1-M/N.
\end{equation*}
Hence,
\begin{equation}
    \label{eq:approxbound1}
    \frac{R_C(M)}{R^\star(M)} \leq \min\{N,K\} \leq 12
\end{equation}
for $\min\{N,K\} \leq 12$.

Assume in the following that $\min\{N,K\} \geq 13$. We consider the
three cases $0 \leq M \leq 1.1\max\{1,N/K\}$, $1.1\max\{1,N/K\} < M
\leq 0.092N$, and $0.092N < M \leq N$ separately.
Assume first that $0 \leq M \leq 1.1\max\{1,N/K\}$. We have
\begin{equation}
    \label{eq:approx1}
    R_C(M) \leq R_C(0) \leq \min\{N,K\}
\end{equation}
by Theorem~\ref{thm:achievability}. By Theorem~\ref{thm:converse}, and
using that $\floor{N/s}\geq N/s-1$,
\begin{equation*}
    R^\star(M) \geq s-\frac{s^2}{1-s/N}\frac{M}{N}.
\end{equation*}
Setting $s=\floor{0.275\min\{N,K\}}\in\{1,\dots,\min\{N,K\}\}$, we
obtain for $M \leq 1.1\max\{1,N/K\}$
\begin{align}
    \label{eq:approx2}
    R^\star(M) 
    & \geq R^\star(1.1\max\{1,N/K\}) \notag\\
    & \geq \floor{0.275\min\{N,K\}}
    - \frac{(\floor{0.275\min\{N,K\}})^2}{1-\floor{0.275\min\{N,K\}}/N}
    \frac{1.1\max\{1,N/K\}}{N} \notag\\
    & \geq \min\{N,K\}
    \biggl(0.275 -\frac{1}{\min\{N,K\}}
    -\frac{1.1\cdot 0.275^2}{1-0.275\min\{1,K/N\}}\biggr) \notag\\
    & \geq \min\{N,K\}\Bigl(0.275-\frac{1}{13}
    - \frac{1.1\cdot 0.275^2}{1-0.275} \Bigr) \notag\\
    & \geq \frac{1}{12}\min\{N,K\},
\end{align}
where we have used the assumption that $\min\{N,K\} \geq 13$.
Combining \eqref{eq:approx1} and \eqref{eq:approx2} yields
\begin{equation}
    \label{eq:approxbound2}
    \frac{R_C(M)}{R^\star(M)} \leq 12
\end{equation}
for $0 \leq M \leq 1.1\max\{1,N/K\}$ and $\min\{N,K\} \geq 13$.

Assume next that $1.1\max\{1,N/K\} < M \leq 0.092N$. Let $\tilde{M}$ be the
largest multiple of $N/K$ less than or equal to $M$, so that
\begin{equation*}
    0 \leq M-N/K \leq \tilde{M} \leq M.
\end{equation*}
Then, by Theorem~\ref{thm:achievability},
\begin{align}
    \label{eq:approx3}
    R_C(M) 
    & \leq R_C(\tilde{M}) \notag\\
    & \leq \frac{K(1-\tilde{M}/N)}{1+K\tilde{M}/N} \notag\\
    & \leq \frac{K(1-M/N+1/K)}{1+KM/N-1} \notag\\
    & \leq N/M,
\end{align}
where we have used that $M/N \geq 1/K$ in the last inequality.
Setting $s=\floor{0.3N/M}\in\{1,\dots,\min\{N,K\}\}$ in
Theorem~\ref{thm:converse}, we obtain 
\begin{align}
    \label{eq:approx4}
    R^\star(M)
    & \geq s-\frac{s^2}{N-s}M \notag\\
    & \geq \frac{0.3N}{M}-1-\frac{0.3^2N^2/M^2}{N-0.3N/M}M \notag\\
    & \geq \frac{N}{M}\Bigl(0.3-0.092-\frac{0.3^2}{1-0.3/1.1}\Bigr) \notag\\
    & \geq \frac{N}{12M}.
\end{align}
Combining \eqref{eq:approx3} and \eqref{eq:approx4} yields
\begin{equation}
    \label{eq:approxbound3}
    \frac{R_C(M)}{R^\star(M)}
    \leq 12
\end{equation}
for $1.1\max\{1,N/K\} < M \leq 0.092N$. 

Finally, assume $0.092N < M \leq N$. Let $\tilde{M}$ be the largest multiple of
$N/K$ less than or equal to $0.092N$, so that
\begin{equation*}
    0 \leq 0.092N-N/K \leq \tilde{M} \leq 0.092N.
\end{equation*}
Then, using convexity of $R_C(\cdot)$ and that $R_C(N) = 0$, 
\begin{align*}
    R_C(M) 
    & \leq \frac{R_C(\tilde{M})}{1-\tilde{M}/N}(1-M/N) \\
    & \leq \frac{1}{1-0.092}R_C(\tilde{M})(1-M/N),
\end{align*}
where we have used that $\tilde{M} \leq 0.092N$.
Now, by Theorem~\ref{thm:achievability},
\begin{align*}
    R_C(\tilde{M}) 
    & \leq \frac{1}{\tilde{M}/N+1/K} \\
    & \leq \frac{1}{0.092-1/K+1/K} \\
    & = \frac{1}{0.092}.
\end{align*}
Hence,
\begin{align}
    \label{eq:approx5}
    R_C(M) 
    & \leq \frac{1}{0.092(1-0.092)}(1-M/N) \notag\\
    & \leq 12(1-M/N).
\end{align}
Setting $s=1$ in Theorem~\ref{thm:converse}, we obtain 
\begin{equation}
    \label{eq:approx6}
    R^\star(M) \geq 1-M/N.
\end{equation}
Combining \eqref{eq:approx5} and \eqref{eq:approx6} yields
\begin{equation}
    \label{eq:approxbound4}
    \frac{R_C(M)}{R^\star(M)} \leq 12
\end{equation}
for $0.092N < M \leq N$.

Combining \eqref{eq:approxbound1}, \eqref{eq:approxbound2},
\eqref{eq:approxbound3}, and \eqref{eq:approxbound4} yields
\begin{equation*}
    \frac{R_C(M)}{R^\star(M)} \leq 12
\end{equation*}
for all $N,K$ and all $0 \leq M \leq N$.\hfill\IEEEQED

\section{Discussion and Directions for Future Work}
\label{sec:discussion}

We now discuss the connection of the caching problem to the index and
network coding problems and list some follow-up work as well as
directions for future research.

\subsection{Connection to Index and Network Coding}
\label{sec:discussion_index}

The caching problem introduced in this paper is related to the index
coding problem~\cite{birk06,bar-yossef11} (or,
equivalently~\cite{effros12}, the network coding
problem~\cite{ahlswede00}). We now explain this connection.

The caching problem consists of a placement phase and a delivery phase.
The most important aspect of this problem is the design of the placement
phase in order to facilitate the delivery phase for any possible user
demands. Now, for \emph{fixed} content placement and for \emph{fixed}
demands, the delivery phase of the caching problem induces a so-called
index coding problem. However, it is important to realize that the
caching problem actually consists of exponentially many parallel such
index coding problems, one for each of the $N^K$ possible user demands.
Furthermore, the index coding problem itself is computationally hard to
solve even only approximately~\cite{langberg11}. 

The main contribution of this paper is to design the content placement
such that each of the exponentially many parallel index coding problems
has simultaneously an efficient and analytical solution. This, in turn,
allows to solve the caching problem within a constant factor of
optimality. The proposed coded caching scheme is based on prefetching
uncoded raw bits and on delivering linearly encoded messages. This
contrasts with the index coding problem, where linear codes are not
sufficient to operate within a bounded factor of optimality (i.e.,
nonlinear codes can offer unbounded gain)~\cite{Blasiak11}.

\subsection{Decentralized Caching}
\label{sec:discussion_decentralized}

The coded caching scheme described in this paper has a placement phase
that is orchestrated by a central server. Crucially for this scheme,
both the number and the identity of the  users in the delivery phase are
already known in the prior placement phase. This is clearly not a
realistic assumption since we usually do not know in the morning which
users will request content in the following evening.  Moreover, if
instead of the synchronized user requests here we have more realistic
asynchronous requests, then users join and leave the system over a
period of several hours during the delivery phase, resulting in a
time-varying number of users. Finally, users may be in different
networks during the two phases, e.g., a mobile connected to a wireless
local area network (e.g.  WiFi) during the placement phase and connected
to a cellular network (e.g. LTE) during the delivery phase.

To deal with these issues, we need a placement phase that is
decentralized. We have developed such a decentralized coded caching
scheme in follow-up work~\cite{maddah-ali13}, in which each user caches
a randomly selected subset of the bits. We show that the rate of this
scheme is within a constant factor of optimal universally for any number
of users $K$. Using this universality property allows to address the
problem of asynchronous user requests and of differing networks during
the two phases.

\subsection{Online Caching}
\label{sec:discussion_online}

The caching problem here has two distinct phases: placement and
delivery. The cache is updated only during the placement phase, but not
during the delivery phase. Many caching systems used in practice use
online cache updates, in which a decision to update the cache is made
during file delivery. One popular update rule is least-recently used
(better known by its abbreviation LRU), in which the least-recently
requested file is evicted from the cache~\cite{sleator85}.

It is hence of interest to develop an online version of the coded
caching algorithm proposed here. We present preliminary results in this
direction in~\cite{pedarsani13}, where we introduce a \emph{coded
least-recently sent} delivery and update rule. For a probabilistic model
of user requests, this update rule is shown to be approximately optimal
in~\cite{pedarsani13}. An open question is to find schemes that have
stronger competitive optimality guarantees for individual sequences of
user requests as in~\cite{sleator85}.

\subsection{Nonuniform File Popularities}
\label{sec:discussion_nonuniform}

Throughout this paper, we have adopted a worst-case definition of rate
with respect to user requests. However, different pieces of content have
usually different popularities (i.e., probabilities of being requested
by the users). In order to capture this effect, the definition of
rate needs to be changed from worst case to expected value. 

We report initial results extending the coded caching approach to this
setting in~\cite{niesen13}. The optimality results there are however
weaker, providing only an approximation to within a factor that is (in
most cases of interest) logarithmic in the number of users. Finding
better schemes for this important setting of nonuniform file
popularities and more precisely quantifying the impact of the file
distribution on the global gain arising from coding is hence of
interest.  Furthermore, the results in~\cite{niesen13} hold only when
all users have the same file popularity distribution. Analyzing the
impact of different per-user file popularities is an open problem. 

\subsection{More General Networks}
\label{sec:discussion_networks}

The discussion in this paper focuses on a basic caching network
consisting of a single shared link. To be relevant in practice, the
results here need to be extended to more general networks.  We report
extensions to tree networks with caches at the leaves as well as initial
results on caches shared among several users in~\cite{maddah-ali13}.
An interesting extension of the coded caching approach proposed in this
paper to device-to-device networks without a central server is reported
in~\cite{ji13}. Adapting coded caching to general networks and analyzing
asymmetric scenarios arising for example from nonuniform cache sizes are
open questions.

\subsection{Sharper Approximations}
\label{sec:discussion_bounds}

The upper and lower bounds on the memory-rate tradeoff derived in this
paper are shown here to be within a factor $12$ from each other.
Sharpening this approximation guarantee is of interest. As mentioned,
numerical simulations suggest that a more careful analysis should be
able to decrease the gap to within a factor of $5$. To go beyond that,
better lower and upper bounds are needed, both of which we know from
Example~\ref{eg:2x2} can be improved (see also the Appendix). Some
further questions in this context are as follows.
\begin{itemize}
    \item \emph{Linear versus Nonlinear Coding:} As pointed out above,
        nonlinear coding schemes can offer unbounded gain in network and
        index coding~\cite{Blasiak11}, whereas for the caching problem
        linear coding schemes are sufficient to achieve optimality
        within a constant factor. This raises the questions if we can
        improve the gap using nonlinear schemes or if there exist
        caching networks for which nonlinear schemes can offer unbounded
        gain.
    \item \emph{Larger Field Size:} All operations in this paper are
        over the binary field. In contrast, in network and index coding
        larger field sizes are useful. The question is thus if larger
        field sizes can improve the performance of coded caching. 
    \item \emph{Coded Content Placement:} Example~\ref{eg:2x2} presents
        a scenario with two users, two files, and cache size $1/2$,
        where coded content placement can improve the achievable rate. While
        this type of coded content placement is not needed for a
        constant-factor approximation of the memory-rate tradeoff, it
        might lead to a smaller constant factor.
    \item \emph{Zero Error versus Vanishing Error:} The problem setting
        as described in Section~\ref{sec:problem} allows for a vanishing
        error probability. However, the proposed achievable scheme has
        zero error probability, while still being approximately optimal.
        Schemes with vanishing error probability or lower bounds making
        explicit use of the zero-error requirement might be used to find
        sharper approximations.
\end{itemize}

\subsection{Implementation Complexity}
\label{sec:discussion_size}

Compared to uncoded caching, the coded caching approach suggested in
this paper imposes additional computational burden on the server and the
users, especially for large values of $K$. One approach to deal with
this burden is to use coded caching only among smaller subgroups of
users. This results in lower computational load at the expense of higher
rates over the shared link. Deriving the fundamental tradeoff between
rate and memory subject to complexity constraints is of great interest.

\appendix

Example~\ref{eg:2x2} in Section~\ref{sec:examples} derives upper and
lower bounds on the memory-rate tradeoff $R^\star(M)$ claimed in
Theorems~\ref{thm:achievability} and~\ref{thm:converse} for $K=N=2$.
While these two bounds are sufficient to characterize $R^\star(M)$ to
within a constant multiplicative gap, as we show in what follows,
neither of the two bounds are tight in general. 

We start by arguing that the achievable scheme can be improved by
focusing on the case $M=1/2$.  As before, we split each file into two
subfiles, i.e.,  $A=(A_1, A_2)$ and $B=(B_1, B_2)$. In the placement
phase, we choose the cache contents as $Z_1=A_1\oplus B_1$ and
$Z_2=A_2\oplus B_2$.  Assume that user one requests file $A$ and user
two requests file $B$. The server can satisfy these requests by
transmitting $(B_1, A_2)$ at rate $R=1$. The other three
possible requests can be satisfied in a similar manner. This shows that
$(M,R)$ pair $(1/2,1)$ is achievable. This new point improves the
boundary of the achievable region from the solid blue to the dotted
black curve in Fig.~\ref{fig:2x2} in Section~\ref{sec:examples}.

We now argue that the lower bound on can also be improved. We have for any
achievable memory-rate pair $(M,R)$,
\begin{align*}
    2RF +2MF  
    & \geq H(X_{(1,2)}, Z_1)+H(X_{(2,1)}, Z_2)  \\
    & = H(X_{(1,2)}, Z_1|W_1)+H(X_{(2,1)}, Z_2|W_1) 
    + I(W_1; X_{(1,2)}, Z_1)+ I(W_1; X_{(2,1)}, Z_2) \\
    & \geq H(X_{(1,2)}, Z_1, X_{(2,1)}, Z_2 | W_1) 
    + I(W_1; X_{(1,2)}, Z_1)+ I(W_1; X_{(2,1)}, Z_2) \\
    & \geq I(W_2; X_{(1,2)}, Z_1, X_{(2,1)}, Z_2 |W_1)  
    +  I(W_1; X_{(1,2)}, Z_1)+ I(W_1; X_{(2,1)}, Z_2).
\end{align*}
Now, since $W_1$ can be decoded from $(X_{(1,2)}, Z_1)$ and also from
$(X_{(2,1)}, Z_2)$, and since $W_2$ can be decoded from $(X_{(1,2)},
Z_1, X_{(2,1)}, Z_2)$, we obtain using Fano's inequality that
\begin{align*}
    I(W_1; X_{(1,2)}, Z_1) & \geq F-\varepsilon F, \\
    I(W_1; X_{(2,1)}, Z_2) & \geq F-\varepsilon F, \\
    I(W_2; X_{(1,2)}, Z_1, X_{(2,1)}, Z_2 |W_1) & \geq F-\varepsilon F
\end{align*}
for any $\varepsilon>0$ as long as $F$ is large enough. Hence,
\begin{align*}
    2RF+2MF \geq 3F- 3\varepsilon F.
\end{align*}
Since $\varepsilon>0$ is arbitrary, this shows that
\begin{equation*}
    R \geq 3/2 - M.
\end{equation*}
Since this is true for any achievable $(M,R)$ pair, this implies that
\begin{equation*}
    R^\star(M) \geq 3/2 - M.
\end{equation*}
This bound, together with the two cut-set bounds derived earlier, proves
that the dotted black curve depicted in Fig.~\ref{fig:2x2} in
Section~\ref{sec:examples} is indeed equal to $R^\star(M)$.

\section*{Acknowledgment}

The authors would like to thank S.~Borst and S.~Kennedy for their
comments on an earlier draft of this paper as well as the reviewers for
their careful reading of the manuscript.

\end{document}